\documentclass[english,eqsecnum,prd,aps,nofootinbib,superscriptaddress,longbibliography,tightenlines,12pt]{revtex4-2}
\usepackage[T1]{fontenc}
\usepackage[latin9]{inputenc}
\usepackage{color}
\usepackage{babel}
\usepackage{mathtools}
\usepackage{amsmath}
\usepackage{amssymb,accents}
\usepackage{mathrsfs}
\usepackage{cancel}
\usepackage{comment}
\usepackage{xcolor}
\usepackage{tensor}
\usepackage{tikz-cd}
\usepackage{tikz}
\usetikzlibrary{shapes,arrows}
\usepackage{booktabs}

\usepackage[unicode=true,pdfusetitle,
bookmarks=true,bookmarksnumbered=false,bookmarksopen=false,
breaklinks=true,allcolors=blue,backref=false,colorlinks=true]
{hyperref}

\definecolor{light-gray}{gray}{0.92}

\begin{document}

\title{Relational Frame Equivalence and Exact Cosmology  in Brans--Dicke Theory}
\author{Yaser Tavakoli}
\email{yaser.tavakoli@ug.edu.pl}
\affiliation{Institute of Theoretical Physics and Astrophysics, Faculty of Mathematics, Physics and Informatics, University of Gdansk, Wita Stwosza 57, 80-308, Gda\'nsk, Poland}

\date{\today}

\begin{abstract}

We apply the canonical clock-sector framework developed in Ref.~\cite{Tavakoli:2026coc} to spatially flat FLRW cosmology in Brans--Dicke  theory. The aim is to obtain an explicit cosmological realization of the frame-adapted relational dynamics and to determine the corresponding cosmological observables in closed form. Starting from the previously developed full ADM theory and restricting it to the homogeneous sector, we obtain the canonical clock momentum as $P_{\Phi}=p_{\phi}-ap_a/2\phi$. This identification allows the Jordan and Einstein-frame reduced systems to be represented on a common canonical phase space.
We then consider a quadratic Jordan-frame potential for which the Einstein-frame potential is constant. The resulting relational FLRW system is exactly integrable. The complete trajectory can be expressed in terms of a single parameter $u$, yielding closed-form expressions for the scale factor, the cosmic-time Hubble rate, the Hubble slow-roll parameter, the number of e-folds, and the scalar equation-of-state parameter. In particular, the exact solution provides an explicit inflationary regime that begins at $u=\operatorname{arctanh}(1/\sqrt{3})$ and transitions toward an asymptotic de Sitter regime.
We finally compare the exact solution with direct numerical integration in the adapted Jordan and Einstein descriptions, and with the naively reduced Jordan system based on the unshifted scalar momentum. The numerical results reproduce the exact trajectory and illustrate how the different reduced canonical representations are related by the clock-dependent transformation implied by the canonical clock-sector construction. 
The analysis provides an exactly solvable laboratory for relational frame equivalence in scalar--tensor cosmology and gives explicit cosmological observables that can be used in extensions to more general potentials and scalar--tensor models.
		
\end{abstract}
\maketitle

\section{Introduction}

Scalar--tensor theories of gravity provide a broad framework for extending general relativity by promoting the gravitational interaction to depend on additional dynamical scalar degrees of freedom \cite{Fujii:2003pa}. Among the simplest and most influential examples is Brans--Dicke (BD) theory, originally introduced as a relativistic realization of Mach's principle in which the effective gravitational coupling is promoted to a dynamical scalar field \cite{Brans:1961sx,Dicke:1961gz}. Because of its conceptual simplicity and its role as a prototype for more general scalar--tensor theories, BD gravity has long served as a useful setting for investigations of gravitation, cosmology, and quantum gravity. Its Jordan- and Einstein-frame formulations have consequently been studied from covariant, Hamiltonian, cosmological, and quantum perspectives \cite{Maeda:1988ab,Faraoni:1998qx,Postma:2014vaa,Pandey:2016unk,Pandey:2016jmv}.
	
A central issue in these studies is the physical equivalence of the Jordan and Einstein frames. At the classical covariant level, the two formulations are related, under suitable assumptions on the conformal factor and scalar-field redefinition, by an invertible field transformation that maps solutions into one another \cite{Maeda:1988ab,Faraoni:1998qx,HerreroValea:2016dnc}. The interpretation of this equivalence becomes more subtle, however, when matter couplings, quantum corrections, or frame-dependent choices of observables are considered. Consequently, the question of frame equivalence is not simply whether two sets of field equations can be transformed into one another, but rather which canonical structures and physical observables are preserved by the transformation \cite{Postma:2014vaa,Kamenshchik:2014waa,Pandey:2016unk,HerreroValea:2016dnc}.

The issue is particularly delicate in the Hamiltonian formulation. General relativity and scalar--tensor theories are constrained systems, so the Hamiltonian generates gauge transformations rather than evolution with respect to a physical time parameter. A relational description can instead be obtained by selecting a dynamical degree of freedom as an internal clock and constructing observables relative to it \cite{Dittrich:2005kc,Rovelli:2001bz}. In a deparametrizable sector, this amounts to solving the Hamiltonian constraint for the momentum conjugate to the chosen clock, thereby obtaining a reduced Hamiltonian that generates evolution with respect to the corresponding relational time \cite{Rovelli:1990ph,Rovelli:2001bz,Dittrich:2004cb,Kuchar:1991qf}.

Relational reduction therefore acts on a complete canonical pair, rather than on a configuration variable alone. This observation is important for conformally related scalar--tensor theories because the Jordan--Einstein transformation acts nontrivially on the canonical momenta. Moreover, the canonical status of the Weyl transformation itself depends on the phase space
under consideration. In particular, Gionti and collaborators have shown that the naive Weyl transformation is not canonical on the extended ADM phase space including lapse and shift variables, while an appropriate canonical transformation can be obtained after suitable gauge fixing and reduction of the gauge sectors \cite{GabrieleGionti:2020drq,Galaverni:2021xhd,GiontiSJ:2023tgx}.
This issue is conceptually distinct from the covariant equivalence of the two frame formulations and must be kept separate from the question addressed here.

The issue directly relevant for the present work is the interplay between canonical frame transformations and relational deparametrization. When the scalar field is used as an internal clock, the Jordan--Einstein transformation leaves its configuration variable unchanged but shifts its conjugate momentum by a contribution from the gravitational sector. Consequently, relational reduction must be formulated in terms of the complete canonical clock pair rather than the clock configuration variable alone. This observation provides the starting point for the cosmological construction developed below.

This issue was addressed systematically in our previous work \cite{Tavakoli:2026coc}, where the canonical clock-sector construction was developed in the full ADM formulation of BD theory. There we showed that the apparent discrepancy between the Jordan and Einstein-frame reduced Hamiltonians originates from using different canonical embeddings of the scalar clock sector. In particular, the scalar configuration is preserved by the Jordan--Einstein transformation, whereas its conjugate momentum acquires a contribution from the gravitational momentum. A frame-adapted canonical chart was constructed in which the clock sector is consistently transformed before relational reduction, yielding canonically equivalent reduced descriptions.

The present construction is complementary to previous Hamiltonian analyses of the Jordan--Einstein transformation. Those studies emphasize that the canonical status of the Weyl transformation depends on the phase space and gauge variables included in the formulation \cite{GabrieleGionti:2020drq,Galaverni:2021xhd,GiontiSJ:2023tgx}. Here we assume the canonical phase space specified in Ref.~\cite{Tavakoli:2026coc} and address a different question: how the previously established frame-adapted clock sector is realized in homogeneous cosmology and how its use determines the resulting relational observables.

We first specialize the canonical theory to spatially flat, homogeneous, and isotropic Friedmann--Lima\^itre--Robertson--Walker (FLRW) configurations. Rather than introducing an independent minisuperspace theory, we obtain the cosmological Hamiltonian by restricting the full ADM phase space to the homogeneous sector. This gives an explicit FLRW realization of the frame-adapted clock sector of Ref.~\cite{Tavakoli:2026coc}. In particular, the homogeneous Jordan-frame scalar momentum contains the gravitational contribution required by the canonical Jordan--Einstein transformation, while the resulting adapted clock momentum coincides with the scalar momentum of the Einstein-frame description. The FLRW model therefore provides a direct setting in which the canonical clock construction can be followed through to an explicit reduced cosmological system.

We then consider a quadratic Jordan-frame potential, $U(\phi)=\frac{1}{2}m^2\phi^2$, for which the corresponding Einstein-frame potential is constant. This choice renders the frame-adapted relational FLRW system exactly integrable and allows the complete reduced trajectory to be expressed in terms of a single parameter $u$. From this trajectory we reconstruct, in closed form, the cosmic-time Hubble rate, the Hubble slow-roll parameter, the number of e-folds, and the scalar equation-of-state parameter. We thereby obtain an exact characterization of the inflationary regime and its future-directed evolution toward an asymptotic de Sitter regime, without invoking a slow-roll approximation. The exact solution also provides an analytic benchmark for the numerical calculations presented later.

The cosmological realization also permits a direct comparison of three canonical descriptions: the frame-adapted Jordan formulation, the Einstein-frame formulation, and the naively reduced Jordan formulation in which the scalar configuration is used as the clock while its original Jordan-frame momentum is retained. The first two descriptions use the common frame-adapted clock sector and therefore give the same reduced dynamics when expressed in the common adapted canonical variables. The naive Jordan reduction instead uses a different canonical embedding of the clock sector and consequently produces a different reduced Hamiltonian and a different representation of the trajectory. We show explicitly how the latter description is related to the adapted one by a clock-dependent canonical transformation of the reduced variables.

We finally use numerical integration to test the exact analytical construction and to compare the different canonical descriptions. Direct numerical integration of the frame-adapted Jordan equations is compared with the exact solution, while an independent integration of the Einstein-frame reduced system provides a numerical comparison of the two adapted descriptions. We also integrate the naively reduced Jordan system and transform its reduced variables to the frame-adapted canonical variables. After this clock-dependent transformation, the resulting trajectory agrees with the frame-adapted one, showing that the apparent differences arise from the choice of canonical representation rather than from a difference in the relational dynamics or reconstructed cosmological observables.

The main contribution of the present paper is the explicit realization of the canonical clock-sector construction in BD FLRW cosmology and its application to an exactly solvable dynamical model. The analysis follows the chain
\begin{align}
\boxed{\substack{\textsf{\small conformal} \\[4pt] \textsf{\small transformation}}} 
\; \rightarrow \;
\boxed{\substack{\textsf{\small clock-sector } \\[4pt] \textsf{\small  identification}}} 
\; \rightarrow \;
\boxed{\substack{\textsf{\small FLRW } \\[4pt] \textsf{\small  reduction}}} 
\; \rightarrow \;
\boxed{\substack{\textsf{\small exact relational } \\[4pt] \textsf{\small trajectory}}} 
\; \rightarrow \;
\boxed{\substack{\textsf{\small cosmological} \\[4pt] \textsf{\small observables}}}
\label{eq:expected-Chain}
\end{align}
The resulting solution provides closed-form expressions for the reduced phase-space trajectory and for the cosmic-time observables reconstructed from it, including the Hubble rate, Hubble slow-roll parameter, number of e-folds, and scalar equation of state.
The comparison of adapted Jordan and Einstein descriptions, together with the analysis of the naive Jordan reduction, then demonstrates explicitly how the canonical clock-sector construction operates in the cosmological setting.

The paper is organized as follows. In Sec.~\ref{sec:canonical-relational}, we recall  the canonical relations from Ref.~\cite{Tavakoli:2026coc} that are needed for the cosmological application, including the Jordan--Einstein transformation and the frame-adapted clock sector.  Section~\ref{sec:FLRW} specializes this structure to spatially flat FLRW cosmology and derives the corresponding reduced variables and Hamiltonian. Section~\ref{sec:exact_cosmology} develops the exactly solvable quadratic model and reconstructs its physical cosmological observables.
Section~\ref{sec:numerical} presents numerical tests of the exact solution and compares the adapted Jordan, Einstein, and naive Jordan descriptions. We conclude in Sec.~\ref{sec:conclusions} with a discussion of the implications for relational cosmology and extensions to more general scalar--tensor models.

\section{Canonical Structure and Relational Reduction}
\label{sec:canonical-relational}
	
We briefly review the canonical structure of BD theory relevant for the relational analysis. Since the full ADM derivation is given in Ref.~\cite{Tavakoli:2026coc}, we focus on the canonical variables, the Jordan--Einstein transformation, and the role of the scalar momentum in relational reduction.
	
\subsection{Jordan--Einstein canonical structure}
\label{sec:canonical-review}

The Jordan-frame BD action is
\begin{equation}
S_{\mathrm{J}} =
\frac{1}{2\kappa^{2}}
\int d^{4}x\sqrt{-g}
\left[
\phi R
-\frac{\omega}{\phi}
g^{\mu\nu}\nabla_{\mu}\phi\nabla_{\nu}\phi
-U(\phi)
\right]
+\text{\footnotesize Boundary term},
\label{eq:jordan-action-review}
\end{equation}
where $\kappa^{2}=8\pi G$, $\omega$ is the BD parameter, and $U(\phi)$ is the scalar potential.
After the ADM decomposition, the canonical variables are
\begin{equation}
	\Gamma_{\rm J} = \{h_{ab},p^{ab};\,\phi,p_{\phi}\},
\end{equation}
with
\begin{align}
		p^{ab}
		&=
		\frac{\sqrt{h}}{2\kappa^{2}}
		\left[
		\phi (K^{ab}-Kh^{ab})
		-\frac{h^{ab}}{N}
		(\dot{\phi}-N^{c}D_{c}\phi)
		\right],
		\label{eq:pab-review}
		\\
		p_{\phi}
		&=
		\frac{\sqrt{h}}{\kappa^{2}}
		\left[
		\frac{\omega}{N\phi}
		(\dot{\phi}-N^{c}D_{c}\phi)
		-K
		\right].
		\label{eq:pphi-review}
\end{align}
The momentum constraint is
\begin{equation}
		\mathcal{H}_{a}
		=
		-2D_{b}p^{b}{}_{a}
		+p_{\phi}D_{a}\phi,
		\label{eq:momentum-review}
\end{equation}
while the Hamiltonian constraint can be written as
\begin{align}
		\mathcal{H}_{N}
		&=
		\frac{2\kappa^{2}}{\sqrt{h}\,\phi}
		\left[
		p^{ab}p_{ab}
		-\frac12p_{h}^{2}
		+
		\frac{\phi^{2}}{4\bar{\omega}}
		\left(
		p_{\phi}-\frac{p_{h}}{\phi}
		\right)^{2}
		\right]
		\nonumber\\
		&\quad
		-\frac{\sqrt{h}}{2\kappa^{2}}
		\left[
		\phi\,{}^{(3)}R
		-2D^{2}\phi
		-\frac{\omega}{\phi}D_{a}\phi D^{a}\phi
		-U(\phi)
		\right],
		\label{eq:hamiltonian-review}
\end{align}
where
\begin{equation}
		\bar{\omega}\equiv\omega+\frac32,
		\qquad
		p_{h}\equiv h_{ab}p^{ab}.
		\label{eq:defs-review}
\end{equation}
	The non-minimal coupling is reflected in the combination
\begin{equation}
		P_{\phi}^{\mathrm{mix}}
		\equiv
		p_{\phi}-\frac{p_{h}}{\phi},
		\label{eq:mix-momentum-review}
\end{equation}
which will be central to the relational reduction.
	
The Einstein-frame variables follow from the Weyl transformation
\begin{equation}
		\tilde g_{\mu\nu}=\phi\,g_{\mu\nu},
		\label{eq:weyl-review}
\end{equation}
	which, in ADM variables, gives
\begin{equation}
		\tilde h_{ab}=\phi h_{ab},
		\qquad
		\tilde N=\sqrt{\phi}\,N,
		\qquad
		\tilde N^{a}=N^{a}.
		\label{eq:adm-weyl-review}
\end{equation}
Keeping $\phi$ as the scalar variable, the induced transformation on the canonical variables is
\begin{align}
		\tilde h_{ab}
		&=\phi h_{ab},
		&
		\tilde p^{ab}
		&=\frac{p^{ab}}{\phi},
		\label{eq:canon-h-review}
		\\
		\tilde\phi
		&=\phi,
		&
		\tilde p_{\phi}
		&=
		p_{\phi}-\frac{p_{h}}{\phi}. 
		\label{eq:canon-pphi-review}
\end{align}
Thus, although the scalar configuration is unchanged, its conjugate momentum is shifted by the gravitational momentum trace. 

The corresponding Hamiltonian and momentum constraints are
\begin{align}
	\tilde{\mathcal H}_N
	&=
	\frac{2\kappa^2}{\sqrt{\tilde h}}
	\left[
	\tilde p^{ab}\tilde p_{ab}
	-
	\frac12\tilde p_{\tilde h}^{\,2}
	+
	\frac{\phi^2}{4\bar\omega}
	\tilde p_\phi^{\,2}
	\right]
	\nonumber\\
	&\qquad
	-
	\frac{\sqrt{\tilde h}}{2\kappa^2}
	\left[
	{}^{(3)}\tilde R
	-
	\frac{\bar\omega}{\phi^2}
	\tilde D_a\phi
	\tilde D^a\phi
	-
	V(\phi)
	\right], \label{eq:HN-Einstein}
	\\[4pt]
	\tilde{\mathcal H}_a
	&=
	-2\tilde D_b\tilde p^b{}_a
	+
	\tilde p_\phi D_a\phi, \label{eq:Ha-Einstein}
\end{align}
where 
\begin{equation}
	V(\phi)
	=
	\frac{U(\phi)}{\phi^2}.
\end{equation}

The momentum transformation in Eq.~\eqref{eq:canon-pphi-review} is the essential ingredient for what follows.
	
\subsection{Relational reduction and apparent frame dependence}
\label{sec:relational-reduction-review}
	
We now recall the part of the relational reduction that is needed for the FLRW application. We use the BD scalar as an internal clock,
\begin{equation}
		\chi(x)\coloneqq\phi(x)-t\approx0,
		\label{eq:gauge-condition-review}
\end{equation}
and impose the corresponding gauge condition. Solving the Hamiltonian constraint for the momentum conjugate to the clock gives the reduced physical Hamiltonian
\begin{align}
	p_{\phi}&=-H_{\rm phys}	\nonumber\\
		&=
		\frac{p_h}{\phi}
		+\sigma\frac{\sqrt{\bar\omega}}{\kappa^2}\sqrt{h}\,\Omega,
		\label{eq:Hphys-compact}
\end{align}
where $\sigma=+1$ and $\sigma=-1$ represent the expanding and contracting branches, respectively, and
\begin{equation}
\Omega
=
\left[
{}^{(3)}R
-\frac{4\kappa^4}{\phi^2h}
\left(
p^{ab}p_{ab}-\frac12p_h^2
\right)
-\frac{U(\phi)}{\phi}
-\frac{1}{\phi}
\left(
2D^2\phi
+\frac{\omega}{\phi}D_a\phi D^a\phi
\right)
\right]^{1/2}.
\label{eq:Omega-review}
\end{equation}
The reduced evolution of a relational observable is therefore
\begin{equation}
\frac{d\mathcal O}{d\phi}
=
\{\mathcal O,H_{\rm phys}\}_{\rm red}.
\label{eq:ReducedEvolution}
\end{equation}

The crucial point is that deparametrization acts on a complete canonical pair: one solves the constraint for the momentum conjugate to the chosen clock, not for the configuration variable alone. Consequently, the relational reduction depends on the canonical realization of the clock sector.
This becomes apparent when the reduction is performed independently in the two conformal frames. In the Jordan frame one solves for $p_\phi$, whereas in the Einstein frame one solves for $\tilde p_\phi$. Since
\begin{equation}
\tilde p_\phi
=
p_\phi-\frac{p_h}{\phi},
\nonumber
\end{equation}
the two reductions isolate different canonical momenta. For the same branch, their physical Hamiltonians are consequently related by 
\begin{equation}
\tilde H_{\rm phys}
=
H_{\rm phys}
+\frac{p_h}{\phi}.
\label{eq:reduced-ham-difference-review}
\end{equation}
The two reduced Hamiltonians consequently differ when they are expressed in terms of the respective unadapted canonical variables. This difference reflects the different canonical realization of the clock sector rather than a difference between the underlying covariant theories.

The appropriate comparison therefore requires transforming the complete clock pair before performing the relational reduction. This is the frame-adapted construction developed in Ref.~\cite{Tavakoli:2026coc}, which we now specialize to homogeneous cosmology.
	
\subsection{Frame-adapted canonical clock sector}
\label{sec:adapted-clock-review}

The canonical construction of Ref.~\cite{Tavakoli:2026coc} identifies the frame-adapted clock pair as
\begin{align}
\Phi&=\phi,
\qquad  
P_{\Phi}
=
p_{\phi}-\frac{p_h}{\phi},
\label{eq:PPhi-adapted}
\end{align}
The remaining canonical variables are chosen as
\begin{align}
Q_{ab}
&=
\phi h_{ab},
\qquad 
P^{ab}
=
\frac{p^{ab}}{\phi}.
\label{eq:Q-adapted} 
\end{align}
In this canonical chart, the clock sector is adapted to the Jordan--Einstein transformation: the scalar configuration is unchanged while its conjugate momentum is precisely the momentum that transforms into the Einstein-frame scalar momentum,
\begin{equation}
		\Phi=\tilde\phi,
		\qquad
		P_{\Phi}=\tilde p_{\phi}.
\label{eq:clock-identification}
\end{equation}
Consequently, relational reduction with respect to the adapted pair $(\Phi, P_\Phi)$ has the same canonical clock sector in the Jordan and Einstein descriptions. The corresponding reduced Hamiltonians are therefore represented on the same reduced canonical variables. We use this construction as established in Ref.~\cite{Tavakoli:2026coc} and now derive its explicit realization in spatially flat FLRW cosmology.

\section{Cosmological realization}
\label{sec:FLRW}
	
We now specialize the canonical framework developed above to a spatially flat, homogeneous, and isotropic FLRW universe. The purpose is not to construct an independent minisuperspace theory, but to obtain the cosmological model by restricting the full ADM formulation to the homogeneous sector. In this way, the canonical variables, constraints, and Jordan--Einstein transformation inherit directly the structure established on the full phase space. In particular, the frame-adapted canonical clock sector has a direct homogeneous realization and provides the natural canonical variables for the relational cosmological dynamics.

\subsection{Jordan and Einstein-frame cosmology}
\label{sec:reduction}
	
For the spatially flat FLRW geometry,
\begin{equation}
ds^2=-N^2(t)dt^2+a^2(t)\delta_{ij}dx^idx^j,
\end{equation}
the homogeneous phase space is
\begin{equation}
\Gamma_{\rm J}^{\rm FLRW}
=
\{a,p_a;\phi,p_\phi\}.
\end{equation}
Rather than deriving the minisuperspace Hamiltonian independently, we restrict the full ADM expressions to homogeneous configurations. Using Eqs.~(\ref{eq:pab-review}) and (\ref{eq:pphi-review}), the canonical momenta are therefore
\begin{align}
p_a
&=
-\frac{3a}{\kappa^2N}
\left(2\phi\dot a+a\dot\phi\right),		\label{eq:p_a_FLRW}
\\
p_\phi
&=
\frac{a^2}{\kappa^2N}
\left(
-3\dot a+\frac{\omega a}{\phi}\dot\phi
\right),
\label{eq:p_phi_FLRW}
\end{align}
with
\begin{equation}
\{a,p_a\}=1,
\qquad
\{\phi,p_\phi\}=1,
\label{eq:Poisson_FLRW}
\end{equation}
and all other elementary brackets vanishing.

The homogeneous Hamiltonian constraint (\ref{eq:hamiltonian-review}) takes the form
\begin{align}
\mathcal H_{\rm J}^{\rm FLRW}
&=
-\frac{\kappa^2}{12a\phi}\,
p_a^2
+
\frac{\kappa^2\phi}{2\bar\omega a^3}
\left(
p_\phi
-
\frac{ap_a}{2\phi}
\right)^2
+
\frac{a^3U(\phi)}{2\kappa^2},
\label{eq:MiniHamiltonianJordan}
\end{align}
where we have replaced $p_h$ in the full ADM--BD theory by
\begin{equation}
    p_h = \frac{ap_a}{2}.
\end{equation}
Consequently, the scalar kinetic sector is governed by
\begin{equation}
P_\Phi
= 
p_\phi-\frac{ap_a}{2\phi},
\label{eq:MiniClock}
\end{equation}
which is precisely the homogeneous restriction of the frame-adapted momentum (\ref{eq:PPhi-adapted}). Thus, the mixed scalar-gravitational momentum structure of the full ADM theory is retained in minisuperspace.
The important consequence is that the scalar clock sector is not represented by $\Phi=\phi$ alone. Its frame-compatible canonical realization is instead 
\begin{equation}
(\Phi,P_\Phi)
=
\left(
\phi,
p_\phi-\frac{ap_a}{2\phi}
\right).
\label{eq:FLRWadaptedclock}
\end{equation}
This provides the explicit FLRW realization of the canonical clock sector used in the relational construction.

The Einstein-frame cosmological variables follow directly by restricting the Jordan--Einstein canonical transformation to the homogeneous sector. From Eqs.~\eqref{eq:canon-h-review} and \eqref{eq:canon-pphi-review} one obtains
\begin{align}
\tilde a
&=
\sqrt{\phi}\,a,
&
\tilde p_{\tilde a}
&=
\frac{p_a}{\sqrt{\phi}},
\label{eq:MiniTransformationMetric}
\\
\tilde\phi
&=
\phi,
&
\tilde p_\phi
&=
p_\phi-\frac{ap_a}{2\phi}.
\label{eq:MiniTransformationScalar}
\end{align}
These variables satisfy
\begin{equation}
\{\tilde a,\tilde p_{\tilde a}\}=1,
\qquad
\{\phi,\tilde p_{\phi}\}=1,
\end{equation}
with all mixed brackets vanishing. Thus, after restricting to the homogeneous sector, the Jordan--Einstein transformation remains canonical on the resulting unreduced phase space, prior to imposing the Hamiltonian constraint or performing the relational reduction.

In these variables the Hamiltonian constraint (\ref{eq:HN-Einstein}) becomes
\begin{align}
\mathcal H_{\rm E}^{\rm FLRW}
&=
-\frac{\kappa^2}{12\tilde a}
\tilde p_{\tilde a}^{\,2}
+
\frac{\kappa^2\tilde\phi^{\,2}}
{2\bar\omega\,\tilde a^3}
\tilde p_\phi^{\,2}
+ \frac{\tilde a^3V(\tilde\phi)}{2\kappa^2}
\approx0.
\label{eq:MiniHamiltonianEinstein}
\end{align}
In the Einstein-frame variables, the mixed Jordan-frame momentum combination is replaced by the canonical scalar momentum,
\begin{align}
\tilde p_\phi
&=
p_\phi-\frac{ap_a}{2\phi} \nonumber \\
& = P_\Phi.
\label{eq:Einsteinmom}
\end{align}
Thus, the Einstein-frame scalar momentum is precisely the homogeneous realization of the frame-adapted canonical clock momentum introduced in the Jordan description. This identification will be used below to compare the adapted Jordan and Einstein reduced systems in a common canonical representation.

\subsection{Comparison of the reduced cosmological theories}
	
We now perform the relational reduction using the BD scalar as the internal clock. At the unreduced level, the Jordan and Einstein-frame FLRW systems are related by the canonical transformation described above. The distinction between the reduced descriptions arises because the constraint is solved for different canonical momenta.
	
If the Jordan-frame constraint is solved for the unshifted scalar momentum \(p_\phi\), one obtains
\begin{equation}
	p_\phi=-H_{\rm phys}^{\rm (J)}.
\end{equation}
By contrast, the Einstein-frame reduction is performed with respect to the canonical momentum
\begin{equation}
		\tilde p_\phi=-H_{\rm phys}^{\rm (E)}.
\end{equation}
Using (\ref{eq:Einsteinmom}) the two reduced Hamiltonians are related by
\begin{equation}
H_{\rm phys}^{\rm (E)}
=
H_{\rm phys}^{\rm (J)}
+
\frac{ap_a}{2\phi}.
\label{eq:MiniShift}
\end{equation}
Thus, the difference between the naively reduced Jordan Hamiltonian and the Einstein-frame Hamiltonian is entirely determined by the shift of the canonical clock momentum. This is the homogeneous realization of the general result established in Ref.~\cite{Tavakoli:2026coc}.

The appropriate comparison is obtained by using the frame-adapted variables
\begin{align}
a_*	=\sqrt{\phi}\,a,
\qquad 
P_{a_*}	&=	\frac{p_a}{\sqrt{\phi}},
\label{eq:FLRWAdaptedMetric}
\end{align}
together with
\begin{equation}
\Phi
=
\phi,
\qquad 
P_\Phi
=
p_\phi-\frac{ap_a}{2\phi}.
\label{eq:FLRWAdaptedClock}  
\end{equation}
These variables satisfy
\begin{equation}
(a_*,P_{a_*};\Phi,P_\Phi)
\equiv
(\tilde a,\tilde p_{\tilde a};\tilde\phi,\tilde p_\phi).
\label{eq:frameAD-equiv}
\end{equation}
Hence the adapted Jordan-frame variables coincide directly with the Einstein-frame canonical variables.

Solving the constraint with respect to the common adapted clock momentum \(P_\Phi\) therefore gives
\begin{equation}
    P_\Phi=-H_{\rm phys}^{*(\rm J)}
=-H_{\rm phys}^{\rm (E)},
\end{equation}
and consequently
\begin{equation}
H_{\rm phys}^{\ast \rm (J)}
=
H_{\rm phys}^{\rm (E)}.
\label{eq:adaptedH=tildeH-FLRW}
\end{equation}
The FLRW model thus provides an explicit realization of the canonical clock-sector construction of Ref.~\cite{Tavakoli:2026coc}. The difference between the naive Jordan and Einstein reductions is therefore a difference in the canonical realization of the clock sector, not a difference in the unreduced FLRW theory.

\subsection{Relational  cosmological observables}
\label{sec:observables}

Once the Hamiltonian constraint has been solved with respect to the common adapted clock \(\Phi\), the reduced evolution of any observable \(\mathcal{O}\) is generated by the corresponding physical Hamiltonian,
\begin{equation}
\frac{d\mathcal O}{d\Phi}
=
\{\mathcal O,H_{\rm phys}\}_{\rm red}.
\label{eq:evolutionHam}
\end{equation}
The scale factor therefore becomes a relational observable,
\begin{equation}
    a=a(\Phi),
\end{equation}
which describes the expansion history directly in terms of the internal clock.

The Hubble rate and other cosmological quantities can subsequently be reconstructed from this relational trajectory. In particular, the number of e-folds between two clock values is
\begin{equation}
N_e(\Phi)
=
\ln\left[
\frac{a(\Phi)}
{a(\Phi_{\rm i})}
\right].
\end{equation}
The Hubble slow-roll parameter may equivalently be written as
\begin{align}
\epsilon_H
=
-\frac{1}{H_t}\frac{dH_t}{dt}
=
-\frac{1}{H_t}
\frac{dH_t}{d\Phi}
\frac{d\Phi}{dt}.
\label{eq:epsilon-H}
\end{align}
Thus the cosmic-time observables are reconstructed from the relational solution once the relation between the internal clock and cosmic time is specified.

For the frame-adapted Jordan and Einstein descriptions, the canonical identification (\ref{eq:frameAD-equiv}) implies that corresponding relational trajectories are represented in the same reduced canonical variables. In particular,
\begin{equation}
a_*(\Phi)=\tilde a(\Phi),
\end{equation}
with analogous identifications for the corresponding reduced momenta and clock variables. Consequently, cosmological observables constructed from the common reduced trajectory are represented consistently in the adapted Jordan and Einstein descriptions.

The explicit expressions for the Hubble rate, slow-roll parameter, number of e-folds, and scalar equation-of-state parameter will be derived from the exact relational solution in the following section.

\section{EXACT RELATIONAL COSMOLOGY}
\label{sec:exact_cosmology}

The preceding sections established the canonical and relational framework required for a consistent comparison of Jordan and Einstein-frame reductions. In particular, the frame-adapted clock sector identifies the canonical momentum conjugate to the BD scalar and leads to a common reduced Hamiltonian in the adapted Jordan and Einstein descriptions. We now use this framework to study the resulting cosmological dynamics in an exactly solvable model.

We consider a spatially flat FLRW cosmology with the quadratic Jordan-frame potential	
\begin{equation}
U(\phi)=\frac{1}{2}m^2\phi^2.
\label{eq:potentialQuad}
\end{equation}
This choice is particularly useful because the corresponding Einstein-frame potential is constant,
\begin{equation}
    V(\Phi)=\frac{1}{2}m^2,
\end{equation}
so that the adapted relational system becomes exactly integrable. The model therefore provides an analytically controlled setting in which the canonical construction can be connected directly to physical cosmological observables.

We proceed in several steps. First, we solve the adapted relational equations exactly and express the reduced trajectory in terms of a single trajectory parameter \(u\). We then reconstruct the cosmic-time Hubble parameter, the Hubble slow-roll parameter, and the number of e-folds. Finally, we express the same dynamics in terms of the effective equation of state and the kinetic-to-potential energy ratio, and characterize the inflationary region directly in the reduced phase space. The numerical analysis in Sec.~\ref{sec:numerical} subsequently provides an independent test of these analytic results and of the canonical mappings between the different reduced descriptions.

\subsection{Quadratic model and exact relational solution}
\label{subsec:exact_relational_solution}

For the quadratic Jordan-frame potential (\ref{eq:potentialQuad}), the Einstein-frame potential  is constant, $V(\Phi)=m^2/2$, and the reduced dynamics generated by the frame-adapted canonical Hamiltonian become exactly integrable. In the adapted canonical variables
\begin{equation}
		\Gamma_\ast = (a_*,P_{a_*};\Phi,P_\Phi),
\end{equation}
the Hamiltonian constraint can be solved for the canonical clock momentum $P_\Phi$. The resulting physical Hamiltonian is
\begin{equation}
		H_{\rm phys}^{*\rm(J)}
		=
		-\sigma\sqrt{\frac{\bar\omega}{6}}
		\frac{a_*}{\Phi}
		\left(
		P_{a_*}^{2}
		-
		\frac{3m^2}{\kappa^4}a_*^4
		\right)^{1/2},
		\label{eq:Hphys_adapted_quad}
	\end{equation}
	where $\sigma=\pm1$ labels the two branches obtained upon solving the Hamiltonian constraint for $P_\Phi$, with  $P_{a_*}=-H_{\rm phys}^{*\rm(J)}$. For the future-directed expanding branch, we choose the sign of the solution consistently with the expanding sector, for which $P_{a_*}<0$. With this choice, the Hamiltonian flow yields a positive physical Hubble rate.

The relational equations generated by Eq.~(\ref{eq:Hphys_adapted_quad}) are
	\begin{align}
		\frac{da_*}{d\Phi}
		&=
		+\sqrt{\frac{\bar\omega}{6}}
		\frac{a_*P_{a_*}}
		{\Phi\sqrt{
				P_{a_*}^{2}
				-\frac{3m^2}{\kappa^4}a_*^4
		}},
		\label{eq:adapted_da}
		\\[4pt]
		\frac{dP_{a_*}}{d\Phi}
		&=
		-\sqrt{\frac{\bar\omega}{6}}
		\frac{
			P_{a_*}^{2}
			-\frac{9m^2}{\kappa^4}a_*^4
		}{
			\Phi\sqrt{
				P_{a_*}^{2}
				-\frac{3m^2}{\kappa^4}a_*^4
			}
		}.
		\label{eq:adapted_dpa}
	\end{align}
	To solve the above system of equations, it is convenient to introduce the dimensionless variables
	\begin{equation}
		x\equiv\Phi,
		\qquad
		y_*\equiv a_*,
		\qquad
		z_*\equiv-\frac{\kappa^2}{m}P_{a_*}.
		\label{eq:dimensionless_variablesAd}
	\end{equation}
	The relational equations then become
	\begin{align}
		\frac{dy_*}{dx}
		&=
		-\sqrt{\frac{\bar\omega}{6}}
		\frac{y_*z_*}
		{x\sqrt{z_*^2-3y_*^4}},
		\label{eq:dimensionless_adapted_y}
		\\[4pt]
		\frac{dz_*}{dx}
		&=
		\sqrt{\frac{\bar\omega}{6}}
		\frac{z_*^2-9y_*^4}
		{x\sqrt{z_*^2-3y_*^4}}.
		\label{eq:dimensionless_adapted_z}
	\end{align}
	The reality condition of the reduced Hamiltonian restricts the trajectory to
	\begin{equation}
		z_*^2\geq 3y_*^4.
		\label{eq:constraint_adapted}
	\end{equation}

For the expanding branch considered here, $y_*>0$ and $z_*>0$, where $z_*$ is defined as the positive magnitude of $-\kappa^2P_{a_*}/m$. We therefore introduce the phase-space ratio
\begin{equation}
r(x)\equiv\frac{z_*(x)}{y_*^2(x)}.
\label{eq:dimless-r}
\end{equation}
The physical region is then characterized by
\begin{equation}
r\geq\sqrt3,
\label{eq:r-region}
\end{equation}
while the relational equations below are regular for $r>\sqrt{3}$. The boundary $r=\sqrt{3}$ is approached asymptotically by the
expanding solution considered below.

The reduced equations (\ref{eq:dimensionless_adapted_y})--(\ref{eq:dimensionless_adapted_z}) consequently take the form
\begin{align}
\frac{1}{y_*}\frac{dy_*}{dx}
&=
-\sqrt{\frac{\bar\omega}{6}}
\frac{1}{x}
\frac{r}{\sqrt{r^2-3}},
\label{eq:rate-yprime-y}
\\[3pt]
\frac{1}{z_*}\frac{dz_*}{dx}
&=
\sqrt{\frac{\bar\omega}{6}}
\frac{1}{x}
\frac{r^2-9}{r\sqrt{r^2-3}}.
\label{eq:rate-zprime-z}
\end{align}
Using the relation
\[\frac{1}{r}\frac{dr}{dx} = \frac{1}{z_\ast}\frac{dz_\ast}{dx}-\frac{2}{y_\ast}\frac{dy_\ast}{dx},\]
together with Eqs.~(\ref{eq:rate-yprime-y}) and
(\ref{eq:rate-zprime-z}), we obtain
\begin{equation}
\frac{dr}{dx}
=
\sqrt{\frac{3\bar\omega}{2}}
\frac{\sqrt{r^2-3}}{x}.
\label{eq:sep-r}
\end{equation}

It is useful to introduce the dimensionless trajectory parameter
\begin{equation}
u=u_{\rm i}
+
\sqrt{\frac{3\bar\omega}{2}}
\ln\!\left(\frac{x}{x_{\rm i}}\right),
\qquad x=\Phi,
\label{eq:us-x-fin}
\end{equation}
which evolves linearly with  $\ln x$.
Then, on the expanding branch $r>\sqrt3$, the solution can be written as
\begin{equation}
r(x)=\sqrt3\cosh u(x),
\label{eq:rx-sol}
\end{equation}
Here, $(x_{\rm i},u_{\rm i})$ specifies the initial point of the trajectory:
\begin{equation}
u_{\rm i}
=
\operatorname{arcosh}\left(
		\frac{r_{\rm i}}{\sqrt3}
		\right),
		\qquad
		r_{\rm i}
		\equiv
		\frac{{z_*}_{\rm i}}{{y_*}_{\rm i}^2}.
\end{equation}
The initial phase-space point fixes the normalization $y_{*\rm i}$ and the initial value \(u_{\rm i}\), with \(u_{\rm i}\) determined by the initial ratio \(r_{\rm i}\). The subsequent evolution of $u$ with the relational clock $x=\Phi$ depends on $\bar\omega$ through Eq.~(\ref{eq:us-x-fin}), while the $u$-parametrized relation (\ref{eq:rx-sol}) is independent of $\bar\omega$.  Thus, for fixed initial data, the reduced trajectory is conveniently parametrized by $u$.

The scale factor follows directly from  Eq.~(\ref{eq:rate-yprime-y}). From Eq.~(\ref{eq:rx-sol}) and the differentiation of Eq.~(\ref{eq:us-x-fin}), we have
\begin{equation}
\sqrt{r^2-3}
=
\sqrt3\sinh u,
\qquad
\frac{du}{dx}
=
\sqrt{\frac{3\bar\omega}{2}}\frac{1}{x},
\end{equation}
where we have used the $u>0$ branch considered here.
Using these relations in Eq.~(\ref{eq:rate-yprime-y}) and integrating, we obtain
\begin{equation}
y_*(x)
=
{y_*}_{\rm i}
\left[
\frac{\sinh u_{\rm i}}
{\sinh u(x)}
\right]^{\frac13}.
\label{eq:ystar-fin}
\end{equation}
The momentum variable follows from $r=z_*/y_*^2$. We therefore obtain
\begin{equation}
z_*(x)
=
\sqrt3\,{y_*}_{\rm i}^2\,
\cosh u(x)\, \left[
\frac{\sinh u_{\rm i}}
{\sinh u(x)}
\right]^{\frac23}.
\label{eq:zstar-fin}
\end{equation}

Thus, the quadratic model admits the exact solution (\ref{eq:us-x-fin}), (\ref{eq:ystar-fin})--(\ref{eq:zstar-fin}), 	which provides a closed-form description of the reduced trajectory. In particular, the reduced trajectory can be parametrized entirely by $u$, with the  phase-space ratio given by
\begin{equation}
		r(u)=\sqrt3\cosh u.
\label{eq:exactPSsol}
\end{equation}
The physical interpretation of this trajectory, including its expansion and inflationary behavior, will be obtained in the following sections after reconstructing the cosmic-time observables.
\medskip

\noindent\textit{Remark.} The trajectory parameter $u$ is logarithmically
related to the BD scalar clock and is proportional, up to an additive
constant, to the canonically normalized Einstein-frame scalar $\varphi$,
defined by
\begin{equation}
\kappa(\varphi-\varphi_{\rm i})
=
\sqrt{\bar\omega}\,
\ln\left(\frac{\Phi}{\Phi_{\rm i}}\right).
\label{eq:canonical_varphi}
\end{equation}
Using Eq.~(\ref{eq:us-x-fin}), this gives
\begin{equation}
u-u_{\rm i}
=
\sqrt{\frac{3}{2}}\,\kappa
(\varphi-\varphi_{\rm i}).
\label{eq:Einstein-Sc}
\end{equation}
Thus, $u$ is a rescaled and shifted version of the canonically normalized
Einstein-frame scalar. Its use here is a convenient parametrization of
the exact reduced trajectory, while the relational clock employed in
the Hamiltonian reduction remains $\Phi$.

\subsection{Physical expansion rate and onset of inflation}
\label{subsec:physical_inflation}
	
The exact relational solution determines the reduced trajectory, but the physical characterization of inflation requires the reconstruction of the expansion rate with respect to cosmic time. We therefore reconstruct the physical Hubble parameter directly from the canonical momentum of the scale factor.

The adapted gravitational momentum reads
\begin{align}
P_{a_*} &= \frac{p_a}{\sqrt{\phi}} = -\frac{3a}{\kappa^2N\sqrt{\phi}}
\big(
2\phi\dot a+a\dot\phi
\big),
\nonumber \\
&= -\frac{6\tilde{a}}{\kappa^2\tilde{N}}\frac{1}{\sqrt{\phi}}
\big(
\phi\dot a+ \tfrac{1}{2}a\dot\phi
\big) = -\frac{6\tilde{a}}{\kappa^2\tilde{N}}
\dot{\tilde{a}}
\nonumber \\
&= -\frac{6a_*}{\kappa^2N_\ast}\frac{da_*}{dt}, 
\end{align}
where,  a dot denotes differentiation with respect to the coordinate time $t$, 
\begin{equation}
	\dot{a}_\ast = \frac{d}{dt}
	(\sqrt{\phi}a)=\frac{1}{\sqrt{\phi}}\left(\frac{1}{2}\dot{\phi}a + \phi\dot{a}\right),
\end{equation} 
and  $N_\ast=\tilde{N}=\sqrt{\phi}N$.
Choosing the proper-time gauge \(N_*=1\), which is the cosmic time associated with the adapted/Einstein-frame metric, the Hubble parameter is
\begin{equation}
H_t \equiv \frac{1}{a_*}\frac{da_*}{dt} = 		-\frac{\kappa^2P_{a_*}}{6a_*^2}.
\label{eq:Hubble-t}
\end{equation}
Thus, the sign of $P_{a_*}$ determines the orientation of the cosmological expansion, as already discussed  under Eq.~(\ref{eq:Hphys_adapted_quad}): for the expanding branch one has $P_{a_*}<0$, so that $H_t>0$.

In terms of the dimensionless phase-space variables introduced above, Eq.~(\ref{eq:Hubble-t}) gives
\begin{equation}
H_t = \frac{m}{6}\frac{z_*}{y_*^2}
= \frac{m}{6}r.
\label{eq:HubbleP-t1}
\end{equation}
Using the exact phase-space solution (\ref{eq:exactPSsol}) the physical Hubble parameter becomes
\begin{equation}
H_t(u)
= \frac{m}{2\sqrt3}\cosh u.
\label{eq:HubbleP-t2}
\end{equation}
This expression is exact and does not rely on a slow-roll approximation.

The relation between the relational parameter $u$ and cosmic time follows directly from the exact scale-factor solution. Since
\begin{equation}
\frac{d\ln a_*}{du}
=
-\frac13\coth u,
\end{equation}
while
\begin{equation}
H_t
=
\frac{d\ln a_*}{dt},
\end{equation}
Eq.~(\ref{eq:HubbleP-t2}) implies 
\begin{equation} 
-\frac13\coth u\frac{du}{dt} = \frac{m}{2\sqrt3}\cosh u. \end{equation} 
Consequently, 
\begin{equation} 
\frac{du}{dt} = -\frac{\sqrt3\,m}{2}\sinh u. 
\label{eq:du-dt} 
\end{equation} 
Therefore, on the positive-$u$ branch, 
\begin{align} 
\frac{du}{dt}<0, \label{eq:u-future} 
\end{align} 
so that the future-directed expanding solution evolves toward decreasing values of $u$. In particular, the cosmological evolution described by the exact solution proceeds from larger positive values of $u$ toward the limiting value $u=0^+$.

We can now determine the physical inflationary regime directly from the cosmic-time expansion rate. The first Hubble slow-roll parameter is defined by 
\begin{equation} 
\epsilon_H \equiv -\frac{\dot{H}_t}{H_t^2}. \label{eq:epsilonH-def} 
\end{equation} 
From Eq.~(\ref{eq:HubbleP-t2}), 
\begin{equation} 
\frac{dH_t}{du} = \frac{m}{2\sqrt3}\sinh u.
\end{equation} 
Combining this result with Eq.~(\ref{eq:du-dt}) gives 
\begin{align} 
\dot{H}_t = \frac{dH_t}{du}\frac{du}{dt} = -\frac{m^2}{4}\sinh^2u. \label{eq:dot H} \end{align} Substitution into Eq.~(\ref{eq:epsilonH-def}), together with Eq.~(\ref{eq:HubbleP-t2}), then yields \begin{align} \epsilon_H(u) = 3\tanh^2u. 
\label{eq:slowrollRel-analyt} 
\end{align}

Eq.~(\ref{eq:slowrollRel-analyt}) determines the inflationary regime exactly, without invoking any slow-roll approximation. Inflation occurs whenever 
\begin{equation} 
\epsilon_H<1.
\end{equation} 
On the positive-$u$ branch this condition is equivalent to \begin{align} 
\tanh u<\frac{1}{\sqrt3}. 
\end{align} 
The transition into the inflationary regime therefore occurs at the finite value 
\begin{equation} 
u_{\rm inf} = \operatorname{arctanh}\frac{1}{\sqrt3}. \label{eq:inflregime} 
\end{equation} 
At this point, the corresponding hyperbolic functions are \begin{equation} 
\sinh u_{\rm inf} = \frac{1}{\sqrt2}, \qquad \cosh u_{\rm inf} 
= \sqrt{\frac32}. 
\label{eq:infl-hyperbolic} 
\end{equation}

Because the future-directed expanding solution satisfies Eq.~(\ref{eq:u-future}), the trajectory crosses the onset of inflation  at $u=u_{\rm inf}$ and subsequently evolves toward smaller values of $u$. The exact accelerated-expansion condition for $u$ is therefore
\begin{equation} 
0<u<u_{\rm inf}, 
\end{equation} 
and the system approaches the limiting regime 
\begin{equation} 
u\rightarrow0^+ \quad\Rightarrow\quad \epsilon_H\rightarrow0, \qquad H_t\rightarrow\frac{m}{2\sqrt3}. \label{eq:infl-asymptotic} 
\end{equation} 
Hence, the exact expanding branch asymptotically approaches a constant-Hubble, potential-dominated regime as $u\rightarrow0^+$. The limiting value $u=0$ is not reached at a finite value of cosmic time; rather, it is approached asymptotically along the future-directed solution. 

The accumulated duration of the inflationary regime, expressed in terms of the number of e-folds, is determined next.

\subsection{Number of e-folds and asymptotic de Sitter regime}
\label{subsec:efolds_inflation}

The exact scale-factor solution $a_*(x)$ also gives the number of e-folds in closed form. For two points $x_{\rm f}<x_{\rm i}$ along the future-directed physical trajectory, we define
\begin{equation}
N_e(x_{\rm i},x_{\rm f})
\equiv
\ln\frac{a_*(x_{\rm f})}{a_*(x_{\rm i})}
=
\ln\frac{y_*(x_{\rm f})}{y_*(x_{\rm i})}.
\label{eq:Ne-i-f}
\end{equation}
Using the solution (\ref{eq:ystar-fin}), this becomes
\begin{equation}
N_e(x_{\rm i},x_{\rm f})
=
\frac13
\ln\left(
\frac{\sinh u_{\rm i}}
{\sinh u_{\rm f}}
\right),
\label{eq:NeRel}
\end{equation}
where
\begin{equation}
		u_{\rm f}
		=
		u_{\rm i}
		+
		\sqrt{\frac{3\bar\omega}{2}}
		\ln\frac{x_{\rm f}}{x_{\rm i}}.
\label{eq:uf}
\end{equation}

In general, since the future-directed branch evolves toward decreasing $u$, Eq.~(\ref{eq:NeRel}) can be written directly in terms of the value of $u$ reached along the trajectory,
\begin{equation}
		N_e(u;u_{\rm i})
		=
		\frac13
		\ln\left(
		\frac{\sinh u_{\rm i}}
		{\sinh u}
		\right).
\label{eq:Ne-u-general}
\end{equation}
This expression is exact and does not rely on a slow-roll approximation.
	
For the inflationary part of the trajectory, the initial value is naturally taken to be the onset value $u_{\rm i}
=
u_{\rm inf}
		=
\operatorname{arctanh}\frac{1}{\sqrt3}$.
The number of e-folds accumulated after the onset of inflation is therefore
\begin{equation}
		N_e^{\rm inf}(u)
		=
		\frac13
		\ln\left(
		\frac{\sinh u_{\rm inf}}
		{\sinh u}
		\right),
		\qquad
		0<u\leq u_{\rm inf}.
\label{eq:onsetInfNe}
\end{equation}
Using the first identity in Eq.~(\ref{eq:infl-hyperbolic}), we obtain
\begin{equation}
		N_e^{\rm inf}(u)
		=
		-\frac13
		\ln\big(\sqrt2\,\sinh u\big).
		\label{eq:Ne-inf-final}
\end{equation}

The asymptotic behavior follows immediately from $\sinh u\simeq u$ for $u\ll u_{\rm inf}$. Hence,
\begin{equation}
		N_e^{\rm inf}(u)
		\simeq
		-\frac13\ln u
		-\frac16\ln2,
		\qquad
		u\ll u_{\rm inf},
\end{equation}
and consequently
\begin{equation}
	N_e^{\rm inf}\rightarrow\infty
		\qquad
		(u\rightarrow0^+).
\label{eq:Ne-divergence}
\end{equation}
The exact future-directed trajectory therefore enters inflation at the finite value $u=u_{\rm inf}$ but does not encounter a finite background endpoint at which inflation terminates. Instead, $u$ decreases monotonically toward the boundary $u=0^+$, while the accumulated number of e-folds grows without bound. Combined with Eq.~(\ref{eq:infl-asymptotic}), this shows that the asymptotic regime is characterized simultaneously by
\begin{equation}
		H_t
		\rightarrow
		\frac{m}{2\sqrt3},
		\qquad
		\epsilon_H\rightarrow0,
		\qquad
		N_e\rightarrow\infty.
\label{eq:asymptotic-inflation-summary}
\end{equation}
	
The energy decomposition associated with this asymptotic regime is examined next.

\subsection{Equation of state and energy decomposition}
\label{subsec:eos_energy}

In the Einstein-frame description, it is convenient to employ the canonically normalized Einstein-frame scalar \(\varphi\), defined in Eq.~(\ref{eq:canonical_varphi}): 
\begin{equation}
\varphi
\equiv 
\sqrt{\frac{\bar\omega}{\kappa^2}}\ln\Phi.
\end{equation}
Then, the inflationary behavior found in previous subsections admits a direct interpretation in terms of the kinetic and potential contributions of this scalar field. The effective energy density and pressure are therefore
\begin{equation}
    \rho=K+\overline V,
    \qquad
    p=K-\overline V,
\end{equation}
where $K=\frac12\dot{\varphi}^2$ is the kinetic energy density and $\overline V(\varphi)$ is the potential energy density.
In fact, by substituting the canonical momenta
\begin{equation}
	P_{a_*}
	=
	-\frac{6a_*}{\kappa^2}\dot a_*,
	\qquad
	P_\Phi
	=
	\frac{\bar\omega a_*^3}{\kappa^2\phi^2}\dot\phi,
\end{equation}
into the  constraint (\ref{eq:MiniHamiltonianEinstein}), we obtain the Friedmann equation
\begin{align}
	\rho &= \frac{3}{\kappa^2}H_t^2
		=
		K+\overline V \nonumber\\
		&=:\frac12\dot\varphi^2+\overline V(\varphi)
		\nonumber\\
		&=
		\frac{\bar\omega}{2\kappa^2\phi^2}\dot\phi^2
		+
		\frac{V(\phi)}{2\kappa^2},
\label{eq:Friedmann}
\end{align}
where, 
\begin{equation}
\overline V(\varphi)
=
\frac{V(\phi)}{2\kappa^2}.
\end{equation}

The corresponding equation-of-state parameter is
\begin{equation}
w
\equiv
\frac{p}{\rho}
=
\frac{K-\overline V}{K+\overline V}.
\label{eq:eos-def}
\end{equation}
For the quadratic model considered here, the canonical Einstein-frame potential $\overline V$ is constant. The Hamiltonian constraint, together with the exact relational solution, gives the kinetic-to-potential ratio $K/\overline V$ and consequently the equation of state $w$ in terms of the trajectory parameter $u$.

The other Friedmann equation (cosmic-time derivative of the Hubble rate) reads
\begin{equation}
\dot{H}_t
=
-\frac{\kappa^2}{2}\dot\varphi^2
=
-\kappa^2K.
\label{eq:Friedmann2}
\end{equation}	
Together with Eqs.~(\ref{eq:HubbleP-t2}) and (\ref{eq:dot H}), we now obtain
\begin{align}
\overline V
&=
\frac{1}{\kappa^2}
\left(
3H_t^2+\dot{H}_t
\right)
=
\frac{m^2}{4\kappa^2},
\end{align}
and
\begin{align}
K
=
-\frac{\dot{H}_t}{\kappa^2}
=
\frac{m^2}{4\kappa^2}\sinh^2u.
\end{align}
Substituting these results into Eq.~(\ref{eq:eos-def}), the equation of state becomes
\begin{equation}
w(u)
=
\frac{\sinh^2u-1}
{\sinh^2u+1}.
\label{eq:EoS1}
\end{equation}

The equation of state therefore evolves monotonically along the future-directed branch. In the large-$u$ regime, the kinetic contribution dominates,
\begin{equation}
		\frac{K}{\overline V}\gg1,
		\qquad
		w\rightarrow1,
\end{equation}
whereas the small-$u$ regime (the vicinity of $u=0^+$) is potential dominated,
\begin{equation}
\frac{K}{\overline V}\rightarrow0,
		\qquad
		w\rightarrow-1.
\label{eq:eos-asymptotic}
\end{equation}
The exact equation of state obtained above has a direct relation with the physical Hubble slow-roll parameter (\ref{eq:slowrollRel-analyt}). Indeed, from Eq.~(\ref{eq:EoS1}),
\begin{equation}
		1+w
		=
		\frac{2\sinh^2u}{1+\sinh^2u}
		=
		2\tanh^2u
		=
		\frac{2}{3}\epsilon_H,
\end{equation}
and hence
\begin{align}
		\epsilon_H
		=
		\frac32(1+w).
\label{eq:epsilon-eos}
\end{align}
Thus, the exact Hubble slow-roll parameter and the equation-of-state description give the same characterization of the trajectory. In particular, the condition for accelerated expansion, $\epsilon_H<1$, is equivalent to
\begin{equation}
	w<-\frac13,
\end{equation}
and therefore identifies the same onset value (\ref{eq:inflregime}).

The energy decomposition thus provides an independent physical interpretation of the exact inflationary trajectory: the onset of inflation corresponds to the transition from kinetic-dominated to potential-dominated expansion, while the subsequent evolution toward $u=0^+$ corresponds to an increasingly potential-dominated regime.

\subsection{Inflationary region in the reduced phase space}
\label{subsec:inflation_phase_space}
	
The preceding results can be expressed directly in terms of the reduced phase-space variables, without referring explicitly to cosmic time. The exact relational solution for $r(u)$ reads
\begin{equation}
r=\frac{z_*}{y_*^2}=\sqrt3\cosh u.
\label{eq:r-phase-space-u}
\end{equation}
On the branch $u\geq0$, the ratio $r$ is bounded from below,
\begin{equation}
	r\geq\sqrt3,
\end{equation}
with the limiting value $r=\sqrt3$ reached as $u\rightarrow0^+$.
	
At the onset of inflation,  using the second identity of (\ref{eq:infl-hyperbolic}) in~(\ref{eq:r-phase-space-u}), gives the value of $r(u)$ at the onset of inflation,
\begin{equation}
		r_{\rm inf}
		=
		\sqrt3\cosh u_{\rm inf}
		=
	\frac{3}{\sqrt2}.
\end{equation}
Consequently, the inflationary portion of the exact trajectory is characterized entirely by the reduced phase-space ratio $r(u)$ as
\begin{equation}
		\sqrt3
		\leq
		r(u)
		<
	\frac{3}{\sqrt2}.
	\label{eq:rel-rx}
\end{equation}
The upper endpoint,
\begin{equation}
		\frac{{z_*}_{\rm inf}}{{y_*}_{\rm inf}^2}
		=
\frac{3}{\sqrt2},
\end{equation}
corresponds to the onset of inflation, where
\begin{equation}
		\epsilon_H=1,
		\qquad
		w=-\frac13.
\end{equation}
As the future-directed trajectory evolves, $u$ decreases and the ratio $z_*/y_*^2$ moves monotonically toward
\begin{equation}
		\frac{z_*(u)}{y_*^2(u)}
		\rightarrow
		\sqrt3,
		\qquad
		u\rightarrow0^+.
\end{equation}
Thus, the asymptotic inflationary regime is represented in the reduced phase space by the approach to the boundary
\begin{equation}
	z_*^2(u)=3y_*^4(u)
\end{equation}
within the physical region.
	
Eq.~(\ref{eq:rel-rx}) provides a direct relational characterization of the inflationary trajectory. The physical condition $\epsilon_H<1$, obtained from the cosmic-time reconstruction, translates into a simple algebraic interval for the reduced phase-space variables. Hence, the exact solution establishes an explicit correspondence between the reduced phase-space trajectory and the physical cosmological evolution.

The same inflationary condition therefore admits three equivalent representations: in terms of the trajectory parameter \(u\), the reduced phase-space ratio \(r\), and the physical quantities \(\epsilon_H\) and \(w\).

\section{Numerical tests of the exact relational dynamics}
\label{sec:numerical}
	
The exact solution of Sec.~\ref{sec:exact_cosmology} provides a stringent benchmark for the numerical analysis. The numerical study is therefore used for three complementary purposes. First, we verify the closed-form adapted relational trajectory. Second, we integrate the Einstein-frame reduced system independently and test the frame-adapted Jordan--Einstein equivalence in the common canonical variables. Third, we integrate the naively reduced Jordan
system. The latter is not introduced as a physically inequivalent model, but as a controlled test of the clock-sector statement: the naive reduced Hamiltonian is a different canonical representative, and equivalence can only be assessed after the clock-dependent transformation of the reduced non-clock variables is included.

We use the dimensionless variables (\ref{eq:dimensionless_variablesAd}). The initial point is chosen to be
\begin{equation}
		x_{\rm i}=8,\qquad y_{*,\rm i}=1,\qquad z_{*,\rm i}=5,
\label{eq:numerical_initial_data_revised}
\end{equation}
so that $z_{*,\rm i}^2>3y_{*,\rm i}^4$. We consider $\bar\omega=10$ and $\bar\omega=100$. The integration is performed 	toward decreasing $x$, following the future-directed branch of the exact solution. The corresponding values of the canonical-scalar parametrization are
\begin{equation}
	u_{\rm i}=\operatorname{arcosh}\left(\frac{5}{\sqrt3}\right),
	\qquad
	u_{\rm end}=0.05.
\end{equation}
The finite endpoint keeps the numerical integration away from the 	square-root boundary at $u=0$, whose asymptotic behavior is already known analytically.

\subsection{Verification of the exact relational trajectory}
\label{subsec:num_exact_verification}
	
\begin{figure}[t]
\centering
\includegraphics[width=0.72\textwidth]{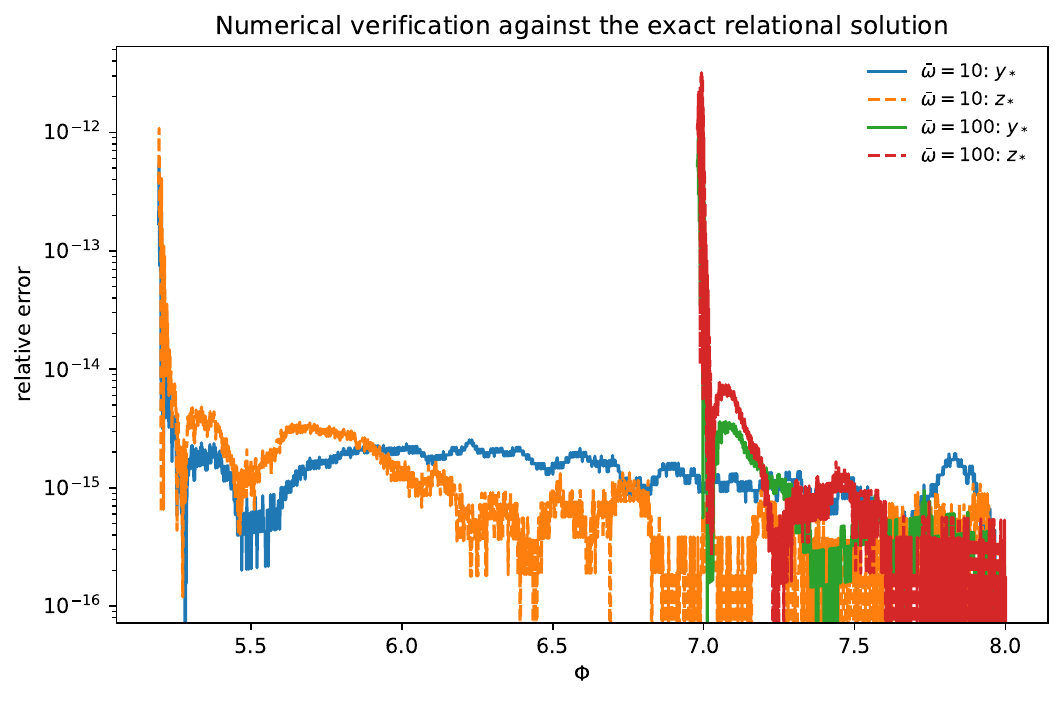}
\caption{Relative deviations of the numerical adapted solution from the exact relational solution for $\bar\omega=10$ and $100$.  The direct numerical integration reproduces the closed-form trajectory with maximum relative deviations below $3.3\times10^{-12}$ over the	displayed interval.}
\label{fig:num-exact-verification}
\end{figure}

We first integrate the adapted relational equations directly in
$(y_*,z_*)$ and evaluate the exact solution at the same values of $x$. We quantify the agreement by
\begin{equation}
\delta_y(x)=\left|\frac{y_*^{\rm num}-y_*^{\rm exact}}
{y_*^{\rm exact}}\right|,\qquad
\delta_z(x)=\left|\frac{z_*^{\rm num}-z_*^{\rm exact}}
{z_*^{\rm exact}}\right|.
\end{equation}
For the integration tolerances used here (cf.~Fig.~\ref{fig:num-exact-verification}), the maximum deviations are
\begin{equation}
	\max\delta_y=\begin{cases}
		4.99\times10^{-13},&\bar\omega=10,\\
		1.70\times10^{-12},&\bar\omega=100,
	\end{cases}
	\qquad
	\max\delta_z=
	\begin{cases}
		9.47\times10^{-13},&\bar\omega=10,\\
		3.21\times10^{-12},&\bar\omega=100.
	\end{cases}
	\label{eq:num_exact_errors_final}
\end{equation}
These deviations show that the numerical integration reproduces the closed-form trajectory with maximum relative deviations below \(3.3\times10^{-12}\) over the
integration interval.

\subsection{Independent Jordan--Einstein numerical equivalence}
\label{subsec:num_frame_equivalence}

In the adapted Jordan description and in the Einstein description, the frame-adapted canonical variables,
\begin{equation}
		a_* =\sqrt{\Phi}\,a,\qquad
		P_{a_*}=\frac{p_a}{\sqrt{\Phi}},\qquad
		P_\Phi=p_\phi-\frac{ap_a}{2\Phi},
\end{equation}
provide the same canonical coordinates on the reduced phase space.
We therefore integrate the two reduced systems independently and
compare $a_*$ and $P_{a_*}$ at fixed $\Phi$.

\begin{figure}[t]
\centering
\includegraphics[width=0.72\textwidth]{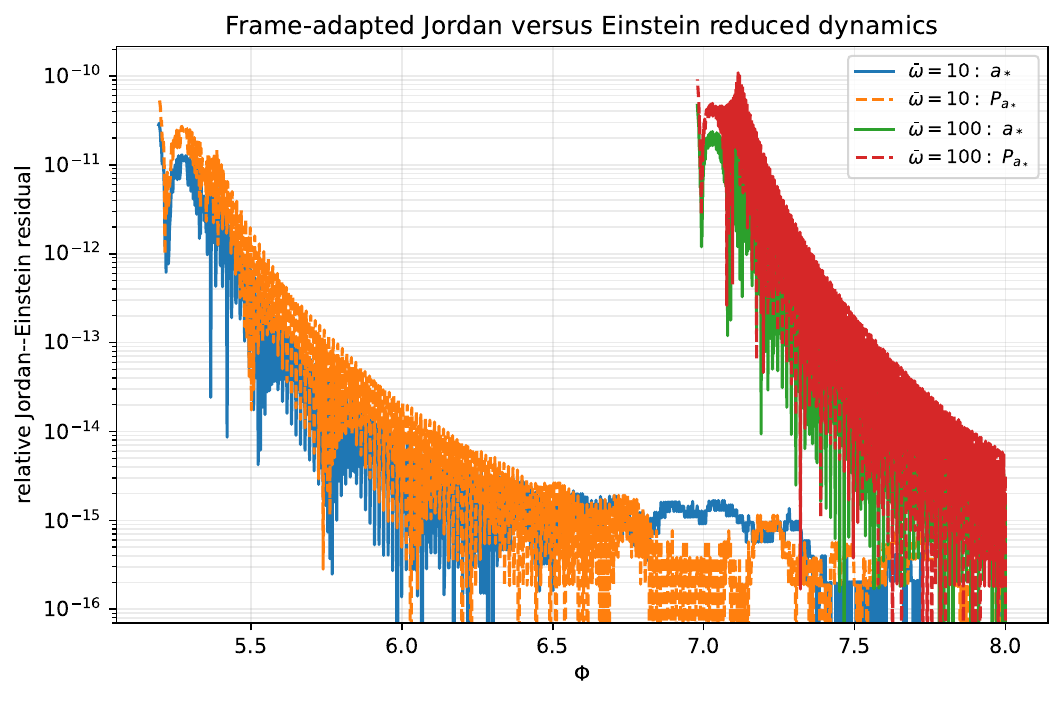}
\caption{Relative residuals between the independently integrated	frame-adapted Jordan and Einstein reduced trajectories, evaluated in the common canonical variables. The residuals remain at the numerical integration level.} 
\label{fig:num-frame-equivalence}
\end{figure}

The relative residuals are
\begin{equation}
\Delta_a(x)=\left|\frac{\tilde a-a_*}{a_*}\right|,\qquad
\Delta_p(x)=\left|\frac{\tilde p_{\tilde a}-P_{a_*}}{P_{a_*}}\right|.
\end{equation}
Their maxima are derived in the plot of Fig.~\ref{fig:num-frame-equivalence} as
\begin{equation}
	\max\Delta_a=\begin{cases}
	1.08\times10^{-10},&\bar\omega=10,\\
	3.34\times10^{-10},&\bar\omega=100,
	\end{cases}
\end{equation}
and
\begin{equation}
\max\Delta_p=
\begin{cases}
	2.11\times10^{-10},&\bar\omega=10,\\
	1.02\times10^{-9},&\bar\omega=100.
\end{cases}
\end{equation}
The larger residuals relative to the exact-solution test simply reflect the independent integration algorithm and tolerances used for the Einstein run. They are numerical residuals rather than a physical frame difference.
	
This calculation provides an independent numerical test of the frame-adapted canonical identification developed in Sec.~\ref{sec:adapted-clock-review}: transforming the complete clock sector before reduction leads to the same reduced trajectory as direct evolution in the Einstein-frame canonical variables, within numerical integration accuracy.

\subsection{Relational clock map and inflationary threshold}
\label{subsec:num_clock_inflation}

The exact solution makes a useful distinction between the phase-space trajectory itself and its parametrization by the BD scalar used as a relational clock. In terms of the trajectory parameter \(u\), the reduced orbit is independent of \(\bar\omega\). In particular, the functions \(y_*(u)\) and \(z_*(u)\) contain no explicit dependence on \(\bar\omega\). Thus, changing \(\bar\omega\) does not change the curve traced in the \((y_*,z_*)\) phase space when the orbit is parameterized by \(u\).

The dependence on \(\bar\omega\) enters through the relation between \(u\) and the relational clock \(\Phi\). From the exact clock map,
\begin{equation}
u(\Phi)
= u_{\rm i}
+
\sqrt{\frac{3\bar\omega}{2}}\,
\ln\left(\frac{\Phi}{\Phi_{\rm i}}\right).
\label{eq:clock_map_exact}
\end{equation}
Equivalently,
\begin{equation}
\Phi(u)
=
\Phi_{\rm i}
\exp\!\left[
\sqrt{\frac{2}{3\bar\omega}}\,
(u-u_{\rm i})
\right].
\label{eq:Phi_of_u_exact}
\end{equation}
Hence, \(\bar\omega\) determines how the same phase-space orbit is parametrized by the clock \(\Phi\). For two different values of \(\bar\omega\), the orbit as a curve in \((y_*,z_*)\) is unchanged when described in terms of \(u\), but the value of \(\Phi\) assigned to each point on that orbit is different. Thus, \(\bar\omega\) changes the clock parametrization of the trajectory rather than deforming the trajectory itself.

\begin{figure}[t]
\centering
\includegraphics[width=0.72\textwidth]{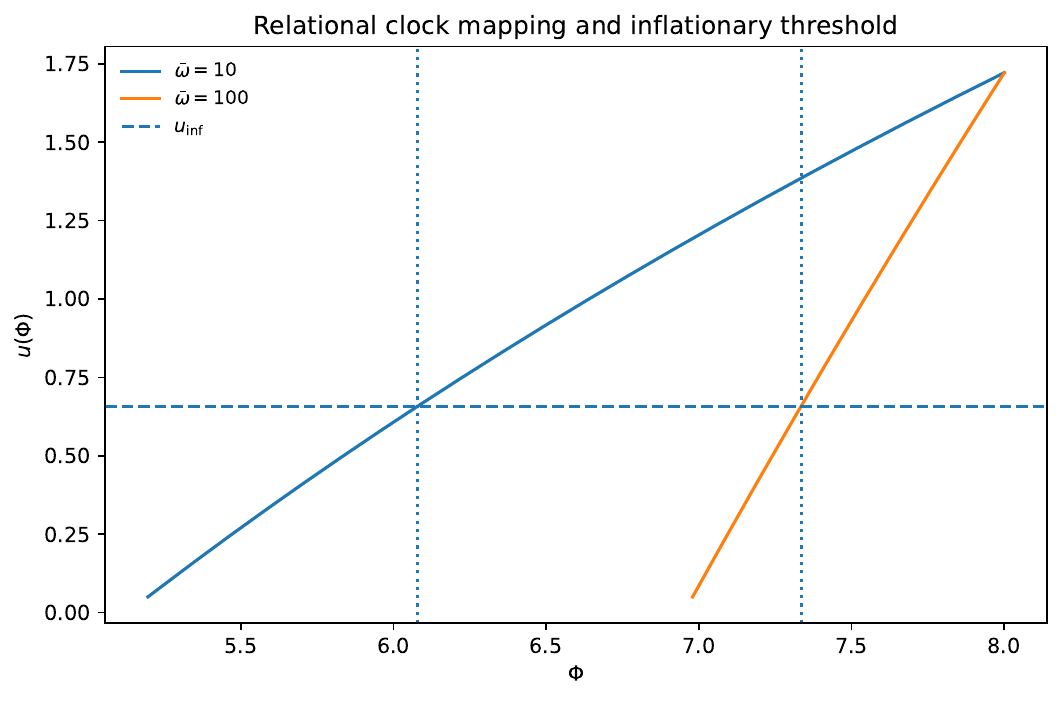}
	
\caption{Exact relational clock map $u(\Phi)$ for $\bar\omega=10$ and $100$.  The curves assign different values of the internal clock $\Phi$ to points on the same exact reduced trajectory, which is parameterized here by $u$.  The inflationary threshold is marked by the horizontal dashed line	$u_{\rm inf}=\operatorname{arctanh}(1/\sqrt3)$; its intersections with the two clock maps occur at $\Phi_{\rm inf}\simeq6.08$ and $7.33$, respectively.}
\label{fig:num-clock-mapping}
\end{figure}

This observation is particularly transparent for the onset of inflation.  The  slow-roll parameter reconstructed from the exact solution is given by Eq.~(\ref{eq:slowrollRel-analyt}). Since the future-directed branch considered here evolves toward $u\rightarrow0^+$,  inflation begins when $\epsilon_H<1$.  The transition to accelerated expansion is therefore determined by
\begin{equation}
	\epsilon_H(u_{\rm inf})=1,
\end{equation}
which gives
\begin{equation}
u_{\rm inf}
=	\operatorname{arctanh}\!\left(\frac{1}{\sqrt3}\right)	\simeq 0.65847895 .
\label{eq:u_inflation_threshold}
\end{equation}
Importantly, $u_{\rm inf}$ is independent of $\bar\omega$; see the horizontal dashed line in Fig.~\ref{fig:num-clock-mapping}. Thus, for
all values of $\bar\omega$, the onset of inflation corresponds to the same point $u=u_{\rm inf}$ on the $u$-parametrized reduced trajectory.
Its coordinate value in the chosen scalar clock, however, depends on $\bar\omega$.

Substituting $u=u_{\rm inf}$ into Eq.~\eqref{eq:Phi_of_u_exact} determines the value of the relational clock $\Phi$ at the onset of inflation:
\begin{equation}
	\Phi_{\rm inf}
	=
	\Phi_{\rm i}
	\exp\!\left[
	\sqrt{\frac{2}{3\bar\omega}}\,
	(u_{\rm inf}-u_{\rm i})
	\right].
	\label{eq:Phi_inflation_threshold}
\end{equation}
For the initial data in Eq.~(\ref{eq:numerical_initial_data_revised}),
\begin{equation}
	u_{\rm i}
	=
	\operatorname{arcosh}\!\left(\frac{5}{\sqrt3}\right)
	\simeq 1.72183119 .
	\label{eq:u_initial_numerical}
\end{equation}
Hence, for the two values considered here,
\begin{equation}
\Phi_{\rm inf}
=
\begin{cases}
		6.07927335, & \bar\omega=10,\\[2mm]
		7.33471969, & \bar\omega=100.
\end{cases}
\label{eq:Phi_inf_numerical}
\end{equation}

The important point is that the inflationary threshold itself is fixed by the condition $\epsilon_H(u_{\rm inf})=1$ and therefore occurs at the same value $u_{\rm inf}\simeq0.65848$ for both values of $\bar\omega$. What changes with $\bar\omega$ is the relation between the trajectory parameter $u$ and the relational clock $\Phi$. Consequently, the same inflationary event is represented by different clock readings,
$\Phi_{\rm inf}=6.07927$ for $\bar\omega=10$ and
$\Phi_{\rm inf}=7.33472$ for $\bar\omega=100$.
Thus, the $\bar\omega$-dependence of $\Phi_{\rm inf}$ should be understood as a change in the clock-coordinate representation of the
event, rather than as a change in the inflationary threshold itself.




The same conclusion can be checked directly using the slow-roll
parameter.  Substituting the clock map
\eqref{eq:clock_map_exact} into Eq.~\eqref{eq:slowrollRel-analyt} gives
the explicit clock representation
\begin{equation}
	\epsilon_H(\Phi;\bar\omega)
	=
	3\tanh^2\!\left[
	u_{\rm i}
	+
	\sqrt{\frac{3\bar\omega}{2}}\,
	\ln\!\left(\frac{\Phi}{\Phi_{\rm i}}\right)
	\right].
	\label{eq:epsilon_phi_exact}
\end{equation}
This expression makes the role of $\bar\omega$ particularly clear: the functional dependence of $\epsilon_H$ on the reduced trajectory parameter $u$ is unchanged, while its representation as a function of the relational clock is stretched or compressed according to
$\bar\omega$.

\begin{figure}[t]
	\centering
	\includegraphics[width=0.72\textwidth]{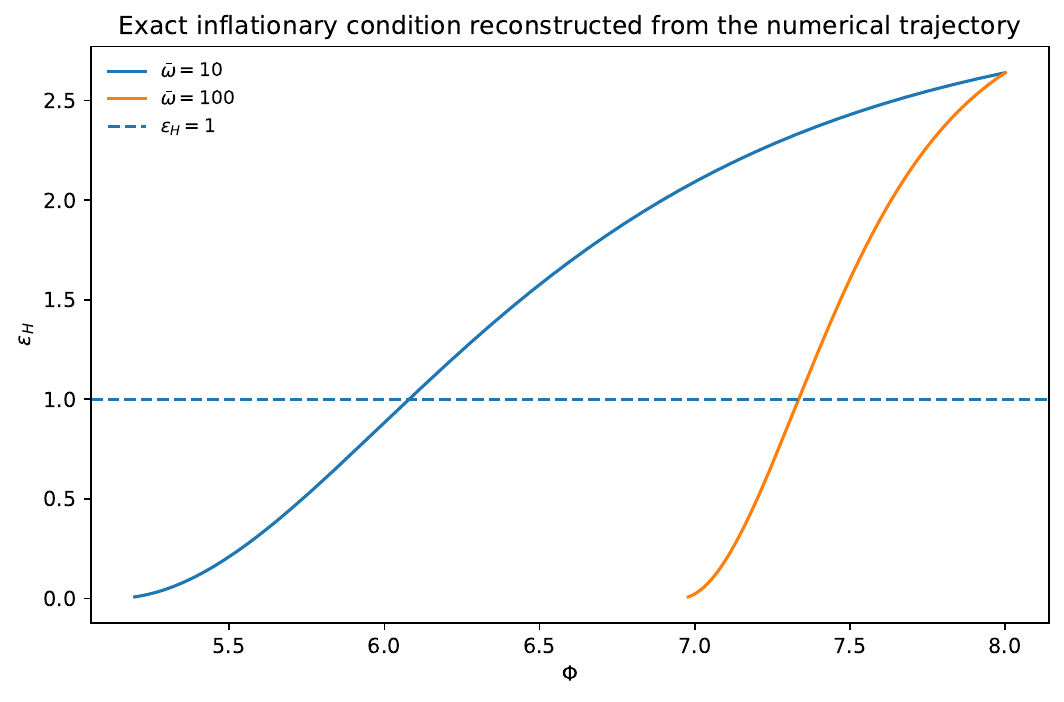}
	\caption{Exact  slow-roll parameter reconstructed in the
		relational clock, $\epsilon_H(\Phi;\bar\omega)$.  The inflationary
		threshold $\epsilon_H=1$ corresponds to the same universal trajectory
		value $u_{\rm inf}$ for both values of $\bar\omega$, while its clock
		coordinate $\Phi_{\rm inf}$ changes according to
		Eq.~\eqref{eq:Phi_inflation_threshold}.  The curves approach
		$\epsilon_H\rightarrow0$ as $u\rightarrow0^+$.}
	\label{fig:num-slowroll}
\end{figure}

Figure~\ref{fig:num-slowroll} displays this exact reconstruction as a function of $\Phi$. Both curves pass through $\epsilon_H=1$ at their respective values of $\Phi_{\rm inf}$ and decrease along the future-directed evolution toward the de Sitter limit $\epsilon_H\rightarrow0$ as $u\rightarrow0^+$.
At the finite numerical endpoint $u_{\rm end}=0.05$, one obtains
\begin{equation}
	\epsilon_H(u_{\rm end})
	\simeq 7.49\times10^{-3}.
	\label{eq:epsilon_endpoint}
\end{equation}
Thus the endpoint used in the numerical integration already lies well
inside the inflationary regime while remaining at a finite distance
from the asymptotic square-root boundary at $u=0$.

The numerical integration provides an independent check of this analytic picture: the directly integrated adapted relational trajectory
reproduces the exact solution to the accuracy reported in Sec.~\ref{subsec:num_exact_verification}, and the reconstructed slow-roll parameter consequently reaches the inflationary threshold at the same universal value $u_{\rm inf}$.  The dependence of
$\Phi_{\rm inf}$ on $\bar\omega$ is therefore a property of the relational clock parametrization, rather than a change in the underlying cosmological trajectory.

\subsection{Naive Jordan reduction and the clock-dependent canonical map}
\label{subsec:num_naive}
	
We next integrate the Jordan reduction in which $\Phi=\phi$ is used as the clock while the unshifted momentum $p_\phi$ is retained. For the expanding branch and the positive variables
\begin{equation}
	y=a,\qquad z=-\frac{\kappa^2}{m}p_a,
\end{equation}
the reduced Hamiltonian, given by solving the constraint (\ref{eq:MiniHamiltonianJordan}), is
\begin{equation}
H_{\rm phys}^{\rm (naive)} = \frac{m}{\kappa^2}\frac{y}{x} 		\left[\sqrt{\frac{\bar\omega}{6}} 		\sqrt{z^2-3x^3y^4}+\frac{z}{2}\right].
\label{eq:Hphys_naive_corrected}
\end{equation}
The corresponding first-order equations are
\begin{align}
\frac{dy}{dx}
&=-\sqrt{\frac{\bar\omega}{6}}
	\frac{y z}{x\sqrt{z^2-3x^3y^4}}
	-\frac{y}{2x},\label{eq:naive_num_alpha_corrected}\\ \nonumber\\
	\frac{dz}{dx}
&=\sqrt{\frac{\bar\omega}{6}}
	\frac{z^2-9x^3y^4}{x\sqrt{z^2-3x^3y^4}}
	+\frac{z}{2x}.
\label{eq:naive_num_q_corrected}
\end{align}
The initial data corresponding to the adapted point in Eq.~\eqref{eq:numerical_initial_data_revised} are
\begin{equation}
y_{\rm i}=\frac1{\sqrt8},\qquad
z_{\rm i}=5\sqrt8.
\end{equation}

\begin{figure}[t]
\centering
\includegraphics[width=0.72\textwidth]{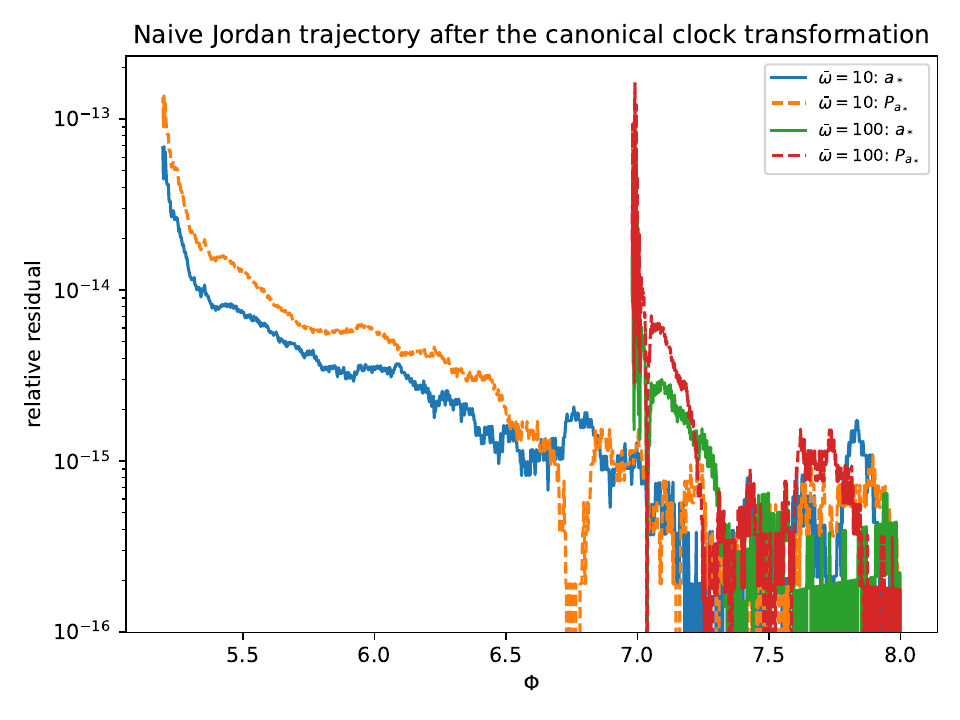}
\caption{Relative residuals between the naive Jordan trajectory after its clock-dependent canonical transformation and the independently integrated adapted trajectory. The residuals remain at the numerical integration level, showing that the two reduced Hamiltonian representatives describe the same relational flow when the reduced observables are mapped consistently.}
\label{fig:num-naive-mapped}
\end{figure}

The crucial point is that the naive and adapted reduced Hamiltonians are not supposed to be literally identical. At fixed clock value $\Phi$, the reduced non-clock variables are related by the time-dependent canonical map [cf. Eq.~(\ref{eq:FLRWAdaptedMetric})]:
\begin{equation}
a_*=\sqrt{\Phi}\,y,
\qquad
P_{a_*}=\frac{p_a}{\sqrt{\Phi}}.
\end{equation}
With the positive momentum variables used above, $z_* =z/\sqrt{\Phi}$ and $y_*=\sqrt{\Phi}\,y$. Applying this map to the numerical naive trajectory gives the same trajectory in the adapted canonical variables, up to the numerical integration accuracy; see Fig.~\ref{fig:num-naive-mapped}.
	
\begin{figure}[t]
\centering
\includegraphics[width=0.72\textwidth]{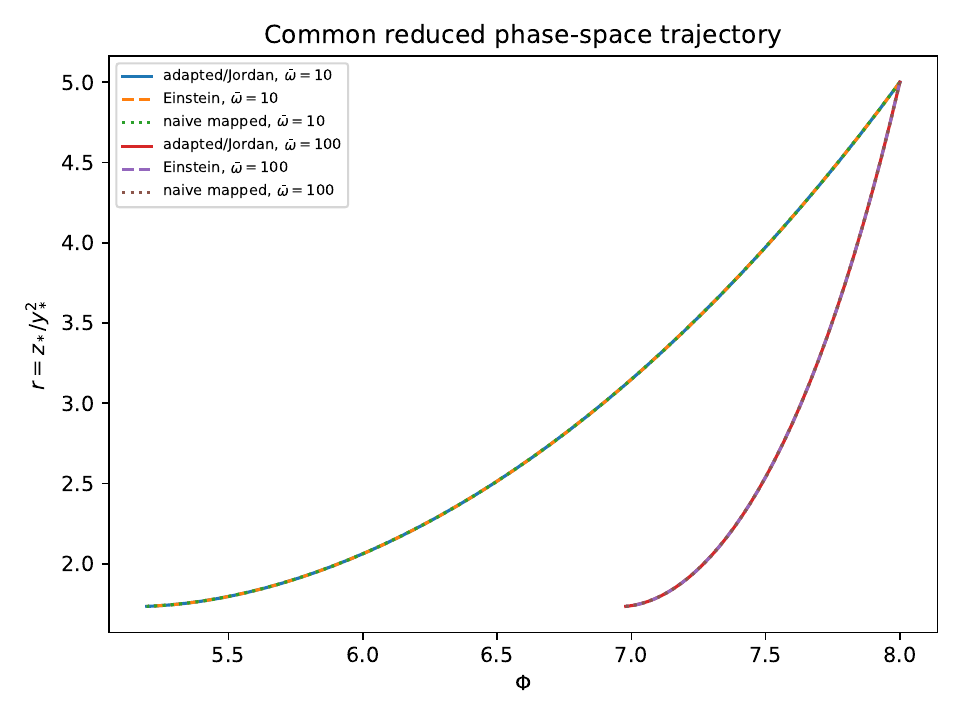}
\caption{The common reduced phase-space ratio $r=z_*/y_*^2$ for the adapted Jordan, Einstein, and canonically mapped naive-Jordan descriptions. The three curves are indistinguishable at the scale of the plot.}
\label{fig:num-phase-space-equivalence}
\end{figure}

The agreement is also visible in the common phase-space ratio,
\begin{equation}
r(\Phi)=\frac{z_*}{y_*^2}=\sqrt3\cosh u(\Phi).
\end{equation}
After expressing the three descriptions in the common adapted canonical variables, the corresponding trajectories coincide within numerical accuracy [see Fig.~\ref{fig:num-phase-space-equivalence}].
Thus the naive reduction does not provide a counterexample to frame equivalence; it provides a counterexample only to the incomplete procedure of comparing untransformed reduced variables across canonical charts.

\subsection{Relational expansion and e-folds}
\label{subsec:num_efolds}
	
For the adapted solution, the exact e-fold count between $\Phi_{\rm i}$ and	$\Phi$ is
\begin{equation}
N_e^*(\Phi)=\ln\frac{a_*(\Phi)}{a_{*,\rm i}}
=\frac13\ln\frac{\sinh u_{\rm i}}{\sinh u(\Phi)}.
\end{equation}
At $u_{\rm end}=0.05$,
\begin{equation}
N_e^*=1.33051024,
\end{equation}
of which $0.88291402$ e-folds are accumulated after the inflationary onset at $u_{\rm inf}$. The mapped naive-Jordan and Einstein descriptions give the same value within numerical precision.

For comparison, the Jordan-frame scale factor $a(\Phi)$ has a different clock-dependent representation. This follows directly from
the conformal relation
\begin{equation}
a_*=\sqrt{\Phi}\,a.
\end{equation}
Consequently,
\begin{equation}
\ln\frac{a(\Phi)}{a_{\rm i}}
=
\ln\frac{a_*(\Phi)}{a_{*,\rm i}}
-\frac12\ln\frac{\Phi}{\Phi_{\rm i}}.
\end{equation}
Thus, the e-fold count defined above,
\[
N_e^*=\ln\frac{a_*}{a_{*,\rm i}},
\]
measures the expansion with respect to the adapted/Einstein-frame
scale factor $a_*$. It should not be identified directly with
$\ln[a(\Phi)/a_{\rm i}]$ in the Jordan frame; the conformal factor
$\sqrt{\Phi}$ must be included when relating the two descriptions.

\begin{figure}[t]
\centering
\includegraphics[width=0.72\textwidth]{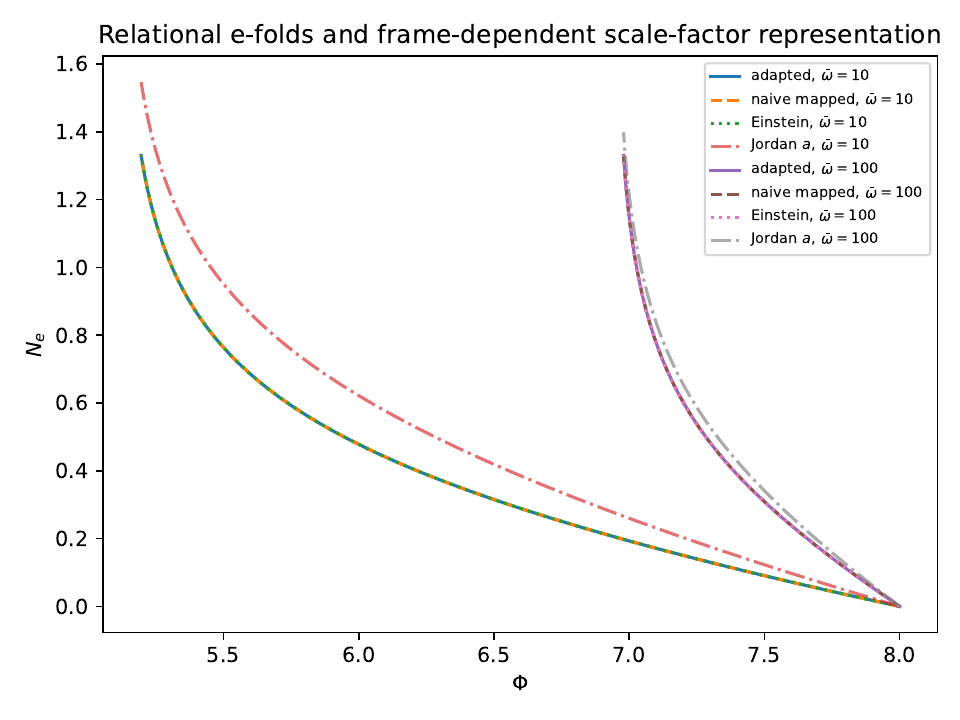}
\caption{Accumulated e-folds as functions of the common relational clock. The adapted, Einstein, and canonically mapped naive-Jordan results coincide. The Jordan-frame scale factor $a$ is shown separately to emphasize that the coordinate representation of the scale factor is frame dependent even when the relationally mapped dynamics are equivalent.}
\label{fig:num-efolds-equivalence}
\end{figure}
	
The numerical results (cf. Fig.~\ref{fig:num-efolds-equivalence}) therefore distinguish two statements that are sometimes conflated. The first is literal equality of Hamiltonian expressions, which is not expected after a time-dependent change of reduced canonical variables. The second is canonical equivalence of the reduced dynamical systems, which is the relevant statement for relational physics. The present calculation verifies the second statement explicitly.
	
\subsection{Numerical summary}

The numerical analysis provides three complementary checks of the analytic construction:

\begin{enumerate}
\item The adapted numerical integration reproduces the exact closed-form trajectory with maximum relative deviations below \(3.3\times10^{-12}\).

\item An independent Einstein-frame integration reproduces the adapted Jordan trajectory in the common canonical variables, with residuals at the \(10^{-10}\)--\(10^{-9}\) level.

\item The naively reduced Jordan trajectory reproduces the same relational orbit after the clock-dependent canonical transformation of the reduced non-clock variables is applied.

\end{enumerate}

Together, these results provide a numerical realization of the analytic canonical construction and illustrate the distinction between equality of Hamiltonian expressions and canonical equivalence of reduced dynamical systems.

\section{Conclusions}
\label{sec:conclusions}

We have applied the canonical clock-sector framework developed in
Ref.~\cite{Tavakoli:2026coc} to spatially flat FLRW cosmology in BD theory. The central point is that relational deparametrization with respect to the BD scalar is defined by a complete canonical
clock sector rather than by the scalar configuration variable alone. Although the scalar configuration is unchanged under the Jordan--Einstein transformation, its canonical momentum is shifted by the gravitational momentum. In the homogeneous reduction this gives
\[P_\Phi
=
p_\phi-\frac{ap_a}{2\phi},\]
which is the FLRW realization of the frame-adapted clock momentum inherited from the canonical construction on the phase space considered here.

Using this adapted clock sector, the Jordan and Einstein descriptions can be represented on the same reduced canonical phase space. The resulting reduced Hamiltonians coincide when expressed in the adapted variables, whereas the naively reduced Jordan Hamiltonian differs because it is obtained by solving for the unshifted momentum \(p_\phi\). This difference does not by itself
signal a physical inequivalence of the two frame descriptions. Rather, the naive and adapted reduced systems are related by a clock-dependent canonical transformation of the reduced non-clock variables. The comparison of relational dynamics therefore requires both the transformation of the
complete clock sector and the corresponding mapping of reduced observables.

Taken together, the analytical and numerical results establish the following chain: on the canonical phase space considered here, the Jordan--Einstein transformation acts canonically; the scalar clock momentum transforms with the gravitational sector; the resulting frame-adapted canonical pair is common to the Jordan and Einstein descriptions; and relational reduction
with respect to this pair produces equivalent reduced Hamiltonian dynamics. The numerical analysis further verifies this construction at the level of the reduced cosmological trajectories and their reconstructed observables.

For the quadratic Jordan-frame potential (\ref{eq:potentialQuad}), the Einstein-frame potential is constant and the adapted relational FLRW system becomes exactly integrable. The complete reduced trajectory is conveniently
encoded by the parameter \(u\), for which
\begin{equation}
r(u)=\frac{z_*}{y_*^2}=\sqrt{3}\cosh u. \nonumber
\end{equation}
The relation between \(u\) and the BD scalar clock depends on \(\bar\omega\). Thus, the \(u\)-parametrized reduced trajectory is independent of \(\bar\omega\), while its parametrization by the chosen internal clock depends on the BD parameter. This separates the reduced trajectory from the particular choice of relational parametrization.

The exact solution also permits the physical cosmological evolution to be reconstructed without invoking a slow-roll approximation. In the
frame-adapted cosmic time, the Hubble parameter and the exact Hubble slow-roll parameter are
\begin{equation}
H_t(u)=\frac{m}{2\sqrt{3}}\cosh u,
\qquad
\epsilon_H(u)=3\tanh^2 u. \nonumber
\end{equation}
Moreover, the corresponding equation of state satisfies
\begin{equation}
w=\tfrac{2}{3}\epsilon_H-1, \nonumber
\end{equation}
providing an equivalent characterization of the exact background evolution. The onset of accelerated expansion occurs at
\begin{equation}
u_{\rm inf}
=
\operatorname{arctanh}\!\big(\tfrac{1}{\sqrt{3}}\big), \nonumber
\end{equation}
and the future-directed expanding branch approaches the asymptotic de Sitter
regime, $H_t\rightarrow\frac{m}{2\sqrt{3}}$,  $\epsilon_H\rightarrow0$, $N_e\rightarrow\infty$,
as \(u\rightarrow0^+\). The corresponding number of e-folds is therefore obtained exactly from the relational solution rather than from a slow-roll approximation.

The numerical analysis provides three complementary tests of the analytical construction. Direct numerical integration of the adapted relational equations reproduces the closed-form trajectory with relative deviations
below \(3.3\times10^{-12}\). An independent numerical integration of the Einstein-frame reduced system reproduces the same trajectory when the solutions are expressed in the common frame-adapted canonical variables. Finally, the naively reduced Jordan system is mapped onto the same reduced trajectory after the clock-dependent canonical transformation of the reduced non-clock variables is applied. These numerical results verify the analytical frame-identification at the level of the reduced cosmological dynamics and provide an explicit consistency check of the corresponding relational observables.

The significance of the FLRW example is therefore primarily methodological. It provides an explicit and exactly solvable realization of how relational
observables should be compared across conformally related canonical descriptions: the complete canonical clock sector must be transformed
consistently, and reduced observables must be compared only after the induced canonical map has been taken into account. The quadratic model
provides a controlled benchmark in which the complete chain from the canonical transformation and relational reduction to the exact cosmological
observables and their numerical verification can be displayed explicitly. The same construction can be extended to nonconstant Einstein-frame
potentials and to more general scalar--tensor theories, where exact solutions may no longer be available and the relational dynamics will
generally require numerical treatment.

\section*{Acknowledgments}

Y.T. acknowledges the Iran National Elites Foundation (BMN) for support through the Kazemi--Ashtiani grant. This work was carried out within the framework of COST Actions CA18108 (Quantum Gravity Phenomenology in the Multi-Messenger Approach), CA23130 (Bridging High and Low Energies in Search of Quantum Gravity), and CA23115 (Relativistic Quantum Information).	
	
\bibliography{main}
	
\end{document}